\documentclass[aps,twocolumn,showpacs]{revtex4}
\usepackage{graphicx}
\usepackage[all]{xy}
\usepackage{amsmath}
\usepackage{amssymb}
\usepackage{epstopdf}
\usepackage[usenames,dvipsnames]{color}
\usepackage[latin1]{inputenc}
\newcommand{\be}{\begin{equation}}
\newcommand{\ee}{\end{equation}}
\newcommand{\bn}{\begin{eqnarray}}
\newcommand{\en}{\end{eqnarray}}
\newcommand{\bes}{\begin{subequations}}
\newcommand{\ees}{\end{subequations}}

\newcommand{\bb}{\bibitem}
\newcommand{\p}{\partial}
\begin{document}

\title{Fast-roll solutions from two scalar fields inflation}
\author{J. R. L. Santos$^{a}${\footnote{email: joaorafael@df.ufcg.edu.br}} and P. H. R. S. Moraes$^{b}${\footnote{email: moraes.phrs@gmail.com}} }
\affiliation{{\small {
$^a$UFCG - Universidade Federal de Campina Grande - Unidade Acad\^{e}mica de F\'isica,  58429-900 Campina Grande, PB, Brazil.\\
$^{b}$USP - University of S\~ao Paulo - Institute of Astronomy, Geophysics and Atmospheric Sciences, S\~ao Paulo SP, Brazil
}
}}

\begin{abstract}

One common approach for cosmic inflation consists in couple Einstein's gravity with a scalar field, often referred to inflaton field. In order to derive analytic simple scenarios, we usually work in the {\it slow-roll} regime. In such an approximation one considers the scalar field potentials to be nearly flat. It is possible to directly generalize such an approach for hybrid inflationary models, where the inflaton sector is composed of two or more scalar fields. However, the Friedmann equations and the equations of motion for such hybrid models are often hard to be analytically solved.  Another recent path to describe cosmic inflation is through the so-called {\it fast-roll} regime, where one considers exactly flat potentials. Our purpose in this work is to obtain solutions for a hybrid inflaton model in the fast-roll regime. Cosmological interpretations through the behavior of Hubble, slow-roll, and equation of state parameters are also presented.

\end{abstract}

\pacs{98.80.Cq; 04.20.Jb}

\maketitle

\section{Introduction}

Scalar field models have been broadly used to describe the phases for which the universe expansion accelerates, i.e., cosmic inflation, and dark energy eras. Such accelerated regimes might be explained from the dynamics of a scalar field $\phi$, which normally is considered a function of time only. Nevertheless it is not easy to analytically solve the cosmological equations of motion for such a scalar field, and sometimes it is necessary to use approximation methods as the {\it slow-roll} regime \cite{linde,albrecht,liddle}. The slow-roll approximation requires $\ddot{\phi}$ and $\dot{\phi}^{2}$ to be small, so that the dynamics of $\phi$ can be given by

\begin{equation}\label{i1}
3H\dot{\phi}\sim-V_\phi,
\end{equation}  
with $H$ being the Hubble parameter, $V_\phi\equiv dV(\phi)/d\phi$ with $V(\phi)$ being a nearly flat scalar field potential and dots stand for derivatives with respect to time.

The slow-roll approximation has also been applied to two scalar fields quintessence models (see for instance \cite{vernizzi/2006,choi/2007,wands/2002}). In \cite{fujii/2000} a primary approach for two scalar fields models has been applied to explain how to obtain a small but non-vanishing cosmological constant. Note that a cosmological constant can also be the responsible for a late time accelerated Universe \cite{riess/1998,perlmutter/1999,hinshaw/2013}, although such an approach raises some shortcomings, as explored in \cite{clifton/2012} and references therein. However it can not account for the early acceleration, since it does not predict the retrieval of a non-accelerated regime after inflation. Anyway, two scalar fields scenarios have been commonly invoked since they exhibit some desirable features, as explored in \cite{linde/1990} for the so-called {\it hybrid inflationary models} and \cite{kofman/1997,bertolami/1986} for reheating models. 

Furthermore, in \cite{ms/2014} it was applied the extension procedure presented in \cite{bls} to solve the two scalar fields equations of motion, obtaining a cosmological scenario which describes radiation, matter and dark energy dominated eras. However, since such an approach was based on analytically solvable models, there was no need of considering the slow-roll approximation.

One could wonder what would be expected when considering exactly flat potentials instead of nearly flat potentials as those used in the slow-roll approximation. An exactly flat potential naturally yields $V_\phi=0$ so that one has

\begin{equation}\label{i2}
\ddot{\phi}+3H\dot{\phi}\sim0
\end{equation} 
as the Klein-Gordon equation for the inflaton field. Although there is a discreet difference between nearly flat and exactly flat potentials, there are several remarkable differences from one model to another \cite{motohashi/2014}. 

The exactly flat potentials approach, first studied in \cite{kinney/2005}, has been primarily called {\it ultra-slow-roll inflation} \cite{martin/2013}, but since the slow-roll approximation breaks down for $V_\phi=0$, now it is dubbed {\it fast-roll inflation} \cite{motohashi/2014}.

Our proposal in this work is to investigate cosmological solutions for a two scalar fields quintessence model in the fast-roll regime, generalizing the proposal done by Motohashi et al. \cite{motohashi/2014}. We follow the investigation introduced in \cite{bglm}, in which it was shown how to determine first-order differential equations from the second-order ones, related to the dynamics of the field. It is worth noting that in \cite{bglm} such a methodology was also extended for two scalar fields models. 

The paper is organized in the following nutshell: in Section \ref{sec:fri} we introduce the fast-roll inflation mechanism and present a method for solving its equation of motion. In Section \ref{sec:tsfa} we present some generalities about the extension method mentioned above, which is fundamental in the construction of the two-field model. The extension procedure is applied for a fast-roll regime and the solutions for the cosmological parameters are obtained. In Section \ref{sec:cs} we perform the cosmological interpretation of our parameters, and we left section \ref{sec:d} for our final remarks, and perspectives.

\section{Fast-roll Inflation}\label{sec:fri}

Let us consider a minimally coupled scalar field defined by the action

\begin{equation}\label{fri1}
S=\int\,d^4\,x\,\sqrt{-g}\,\left[-\frac{R}{2}+{\cal L}(\phi\,,\p_\mu\,\phi)\right]\,,
\end{equation}
for which we are assuming units such that $4\pi G=c=1$ and the Lagrangian density is given simply by

\begin{equation}\label{fri2}
\mathcal{L}=\frac{\dot{\phi}^2}{2}-V(\phi)\,.
\end{equation}
Eq.(\ref{fri2}) yields the equation of motion

\begin{equation}\label{fri3}
\ddot{\phi}+3\,H\,\dot{\phi}+V_{\phi}=0\,.
\end{equation}
Moreover, when working with Friedmann-Robertson-Walker metric, the Friedmann equations (FEs) are

\begin{equation}\label{fri4}
H^2=\frac{1}{3}\left[\frac{\dot{\phi}^{\,\,2}}{2}+V(\phi)\right]\,,
\end{equation}
\begin{equation}\label{fri5}
2\,\dot{H}=-\dot{\phi}^{\,2}\,,
\end{equation}
for which we have considered $k=0$, i.e., a flat space-time, in accord to recent cosmic microwave background anisotropy observations \cite{hinshaw/2013}, $H=\dot{a}/a$ and $a(t)$ is the scale factor. 

Now recall that in fast-roll regime, $V_\phi$ is negligible, so Eq.(\ref{fri3}) reduces to Eq.(\ref{i2}). In fact, in \cite{martin/2013}, fast-roll inflation was generalized by assuming

\begin{equation}\label{fri6}
\ddot{\phi}+(3+\alpha)H\dot{\phi}=0\,,
\end{equation} 
so that nonzero $\alpha$ implies deviation from fast-roll inflation, i.e., the exactly flat potential corresponds to $\alpha=0$, while the standard slow-roll approximation, to $\alpha\sim-3$.

Furthermore, by minimizing the Einstein-Hilbert action, we obtain the energy-momentum tensor
\be
T_{\mu \nu}=2\,\frac{\partial\,{\cal L}}{\partial\,g^{\mu\nu}}-g_{\mu\nu}\,{\cal L}\,,
\ee
with $T_{\mu\nu}=(\rho,-p,-p,-p)$, where $\rho$ and $p$ are the density and pressure of the Universe, respectively. This scalar field description establishes that 
\be
\rho=\frac{\dot{\phi}^{2}}{2}+V\,; \qquad p=\frac{\dot{\phi}^{2}}{2}-V\,.
\ee
The slow-roll parameters are

\be \label{fri7}
\epsilon_1=-\frac{\dot{H}}{H^2}\,;\qquad \epsilon_n=\frac{\dot{\epsilon}_n}{H\,\epsilon_n}\,,
\ee
and in the slow-roll regime we have $\left|\epsilon_n\right|<<1$, therefore they are negligible for high values of time.

It is possible to determine an analytic solution for Eq.\eqref{fri6}, as it was pointed recently in \cite{motohashi/2014}. Therefore, by defining
\be \label{fri8}
H\equiv \frac{W(\phi)}{2}\,; \qquad \dot{H}=\frac{W_{\,\phi}}{2}\,\dot{\phi}\,,
\ee
the Eqs.\eqref{fri5} and \eqref{fri7} can be rewritten as 
\be  \label{fri9}
\dot{\phi}=-W_{\phi}\,; \qquad W_{\,\phi\,\phi}=(3+\alpha)\,W\,.
\ee
So, we obtain our function $W(\phi)$ to be
\be \label{fri10}
W(\phi)=A\,\cosh\,\left(\sqrt{3+\alpha}\,\phi\right)\,,
\ee
where $A$ is an arbitrary constant. Moreover, from the first-order differential equation for the scalar field we directly obtain
\be \label{fri11}
\phi(t)=\frac{1}{\sqrt{3+\alpha}}\,\log\,\left[\coth\,\left(\frac{3+\alpha}{2}\,A\,t\right)\right]\,.
\ee
Consequently, our analytic cosmological parameters are given by
\be \label{fri12}
H=\frac{A}{2}\,\left[\coth\,\left(\frac{3+\alpha}{2}\,A\,t\right)+\tanh\,\left(\frac{3+\alpha}{2}\,A\,t\right)\right]\,;
\ee
\be \label{fri13}
a=\left\{\sinh\left[(3+\alpha)\,A\,t\right]\right\}^{\,\frac{1}{3+\alpha}}\,;
\ee
\be \label{fri14}
\epsilon_1=\frac{3+\alpha}{\cosh^{\,2}\,\left[(3+\alpha)\,A\,t\right]}\,;
\ee
and
\be \label{fri15}
\epsilon_{2\,n}=-2\,(3+\alpha)\,\tanh\left[(3+\alpha)\,A\,t\right]\,; 
\ee
\be \label{fri16}
\epsilon_{2\,n+1}=\frac{2\,(3+\alpha)}{\cosh^{\,2}\,\left[(3+\alpha)\,A\,t\right]}\,.
\ee
From the slow-roll parameters, we see that 
\be \label{fri17}
2\,\epsilon_1=\epsilon_{2\,n+1}\rightarrow 0\,; \qquad
\epsilon_{2\,n}\rightarrow\, -2\,(3+\alpha)\,,
\ee
for $A\,t\gg 1$, which means that the even slow-roll parameters are not negligible for large values of time, unless $\alpha=-3$ (corresponding to the slow-roll regime). In the next section, we are going to extend this approach to a two field quintessence model.

\section{The two scalar fields approach}\label{sec:tsfa}

\subsection{Generalities}

Let us firstly describe some generalities about the extension method which is going to be fundamental in the construction of the effective two-field model.  A formal introduction of this method can be found in \cite{bls}. Here we start by establishing a connection between scalar field $\phi(t)$ and scalar field $\chi(t)$, mediated by a general function $f$, i.e.
\be \label{tsfa1}
\phi=f(\chi)\,;\qquad \chi=f^{-1}(\phi)\,,
\ee
where $f(\chi)$ is invertible and called  the ``deformation function'' \cite{def_1}. Furthermore the field $\chi(t)$ is associated with another one-field theory model. It is straightforward to determine that
\be \label{tsfa2}
\dot{\phi}=f_{\chi}\,\dot{\chi}\,,
\ee
yielding the first-order differential equations
\be \label{tsfa3}
\dot{\phi}=-W_\phi(\phi)=-f_{\chi}\,W_{\chi}(\chi)\,;\qquad \dot{\chi}=-W_\chi(\chi)\,.
\ee
These relations above can be rearranged in the following form
\be \label{tsfa4}
\phi_\chi=f_\chi= \frac{W_\phi(\chi)}{W_\chi(\chi)}\,,
\ee
which has essentially the same form of the first-order differential equation related to the orbit of an effective two-field model \cite{bls,ms/2014}. 

By making $W_\phi(\phi)=W_\phi(\chi)=W_\phi(\phi,\chi)$, $W_\chi(\chi)=W_\chi(\phi)=W_\chi(\phi,\chi)$ and $g(\phi)=g(\chi)=g(\phi,\chi)$ and respecting the constraints $a_1+a_2+a_3=1$, $b_1+b_2+b_3=1$ and $c_1+c_2+c_3=0$, the extension method consists in applying the deformation function to rewrite Eq.\eqref{tsfa4} in the following equivalent form, which is a first-order differential equation related to the orbit between the fields $\phi$ and $\chi$:
\begin{widetext}  
\be \label{tsfa5}
\phi_\chi=\frac{W_\phi}{W_\chi}\equiv\frac{a_1\,W_\phi(\chi)+a_2\,W_\phi(\phi,\chi)+a_3\,W_\phi(\phi)+c_1\,g(\chi)+c_2\,g(\phi,\chi)+c_3\,g(\phi)}{b_1\,W_\chi(\chi)+b_2\,W_\chi(\phi,\chi)+b_3\,W_\chi(\phi)}\,.
\ee
\end{widetext}
Besides, the functions $W_\phi$ and $W_\chi$ have the property
\be \label{tsfa6}
W_{\phi\,\chi}=W_{\chi\,\phi}\,.
\ee
From \eqref{tsfa5} we can identify $W_\phi$ as
\bn \label{tsfa7}
W_\phi &=& a_1\,W_\phi(\chi)+a_2\,W_\phi(\phi,\chi)\\ \nonumber
&&
+a_3\,W_\phi(\phi)+c_1\,g(\chi)+c_2\,g(\phi,\chi)+c_3\,g(\phi)\,, \nonumber
\en
and $W_\chi$ as
\be \label{tsfa8}
W_\chi=b_1\,W_\chi(\chi)+b_2\,W_\chi(\phi,\chi)+b_3\,W_\chi(\phi)\,.
\ee

Therefore, by applying \eqref{tsfa6} we find the second constraint relation
\bn \label{tsfa9}
&&
b_2\,W_{\chi\,\phi}(\phi,\chi)+b_3\,W_{\chi\,\phi}(\phi)=a_1\,W_{\phi\,\chi}(\chi) \\ \nonumber
&&
+a_2\,W_{\phi\,\chi}(\phi,\chi)+c_1\,g_{\chi}(\chi)+c_2\,g_{\chi}(\phi,\chi)\,,  \nonumber
\en
which we use to calculate the function $g$.

It is known that the Hubble parameter and the FE for a two scalar fields model are
\be \label{tsfa10}
H = \frac{W(\phi,\chi)}{2}\,; \qquad H^2=\frac{1}{3}\,\left[\frac{\dot{\phi}^{\,2}}{2}+\frac{\dot{\chi}^{\,2}}{2}+V(\phi,\chi) \right]\,,
\ee
and we also have the following relations for the density and pressure of the Universe:
\be
\rho=\frac{\dot{\phi}^{2}}{2}+\frac{\dot{\chi}^{2}}{2}+V(\phi,\chi)\,; \,\,\,\, p =\frac{\dot{\phi}^{2}}{2}+\frac{\dot{\chi}^{2}}{2}-V(\phi,\chi)\,.
\ee
Those can be used to compute the so-called equation of state (EoS) parameter, which is defined as
\be
\omega=\frac{p}{\rho}=\frac{\dot{\phi}^2+\dot{\chi}^2-2\,V}{\dot{\phi}^2+\dot{\chi} ^2+2\,V}\,.
\ee
Moreover, the scale factor and the slow-roll parameters still obey the relations presented in Eqs.\eqref{fri5} and \eqref{fri7}. 

In the next subsection, the extension method is applied for the fast-roll regime. Such a procedure yields the achievement of some cosmological parameter solutions, which makes possible to check if the model indeed describes the inflationary era.
\subsection{The Extension Method for Fast-Roll Regime}\label{ss:ema}

Once our idea is to construct a two-field model in the fast-roll regime, we are going to deal with a simple deformation function given by
\be \label{ema1}
\phi(t)=\chi(t)=\frac{1}{\sqrt{3+\alpha}}\,\log\,\left[\coth\,\left(\frac{3+\alpha}{2}\,A\,t\right)\right]\,.
\ee
As we have seen in the previous section, these fields are related to
\be \label{ema2}
W_\phi(\chi)=W_\chi(\chi)=A\,\sqrt{3+\alpha}\,\sinh\,\left(\sqrt{3+\alpha}\,\chi\right)\,.
\ee
Now, we can use the deformation function to express such an equation as
\begin{widetext}
\be \label{ema3}
W_{\phi}(\phi)=A\,\sqrt{3+\alpha}\,\sinh\,\left(\sqrt{3+\alpha}\,\phi\right)\,; \qquad 
W_{\phi}(\phi,\chi)=2\,A\,\sqrt{3+\alpha}\,\sinh\,\left(\frac{\sqrt{3+\alpha}}{2}\,\chi\right)\,\cosh\,\left(\frac{\sqrt{3+\alpha}}{2}\,\phi\right)\,,
\ee
\be \label{ema4}
W_{\chi}(\phi)=A\,\sqrt{3+\alpha}\,\sinh\,\left(\sqrt{3+\alpha}\,\phi\right)\,;
\qquad
W_{\chi}(\phi,\chi)=2\,A\,\sqrt{3+\alpha}\,\sinh\,\left(\frac{\sqrt{3+\alpha}}{2}\,\chi\right)\,\cosh\,\left(\frac{\sqrt{3+\alpha}}{2}\,\phi\right)\,.
\ee
If we choose $c_1=0$, Eq.\eqref{tsfa9} implies in the constraint 
\bn \label{ema5}
c_2\,g_{\chi}(\phi,\chi)&=& b_2\,A\,(3+\alpha)\,\sinh\,\left(\frac{\sqrt{3+\alpha}}{2}\,\chi\right)\,\sinh\,\left(\frac{\sqrt{3+\alpha}}{2}\,\phi\right) \\ \nonumber
&-&
a_2\,A\,(3+\alpha)\,\cosh\,\left(\frac{\sqrt{3+\alpha}}{2}\,\chi\right)\,\cosh\,\left(\frac{\sqrt{3+\alpha}}{2}\,\phi\right) \\ \nonumber
&+&
b_3\,A\,(3+\alpha)\,\cosh\,\left(\sqrt{3+\alpha}\,\phi\right)-a_1\,A\,(3+\alpha)\,\cosh\,\left(\sqrt{3+\alpha}\,\chi\right)\,,
\en
therefore, by integrating it with respect to the field $\chi$ we obtain
\bn \label{ema6}
c_2\,g(\phi,\chi)&=&2\,b_2\,A\,\sqrt{3+\alpha}\,\cosh\,\left(\frac{\sqrt{3+\alpha}}{2}\,\chi\right)\,\sinh\,\left(\frac{\sqrt{3+\alpha}}{2}\,\phi\right) \\ \nonumber
&-&
2\,a_2\,A\,\sqrt{3+\alpha}\,\sinh\,\left(\frac{\sqrt{3+\alpha}}{2}\,\chi\right)\,\cosh\,\left(\frac{\sqrt{3+\alpha}}{2}\,\phi\right) \\ \nonumber
&+&b_3\,A\,(3+\alpha)\,\cosh\,\left(\sqrt{3+\alpha}\,\phi\right)\,\chi-a_1\,A\,\sqrt{3+\alpha}\,\sinh\,\left(\sqrt{3+\alpha}\,\chi\right)\,,
\en
and the deformation function presented in \eqref{ema1} allows us to rewrite the previous equation as
\be \label{ema7}
c_2\,g(\phi) = (b_2-a_1-a_2)\,A\,\sqrt{3+\alpha}\,\sinh\,\left(\sqrt{3+\alpha}\,\phi\right)+b_3\,A\,(3+\alpha)\,\cosh\,\left(\sqrt{3+\alpha}\,\phi\right)\,\phi\,.
\ee
Thus the effective $W_{\phi}$ and $W_{\chi}$ are such that
\bn \label{ema8}
W_{\phi}&=&A\,\sqrt{3+\alpha}\,\sinh\,\left(\sqrt{3+\alpha}\,\phi\right)+\,b_2\,A\,\sqrt{3+\alpha} \bigg[2\,\cosh\,\left(\frac{\sqrt{3+\alpha}}{2}\,\chi\right)\,\sinh\,\left(\frac{\sqrt{3+\alpha}}{2}\,\phi\right) \\ \nonumber
&&
-\sinh\,\left(\sqrt{3+\alpha}\,\phi\right)\bigg]
+b_3\,A\,(3+\alpha)\,\left[\cosh\,\left(\sqrt{3+\alpha}\,\phi\right)\,\chi-\cosh\,\left(\sqrt{3+\alpha}\,\phi\right)\,\phi\right]\,;
\en
\bn \label{ema9}
W_\chi&=&b_1\,A\,\sqrt{3+\alpha}\,\sinh\,\left(\sqrt{3+\alpha}\,\chi\right) \\ \nonumber
&&
+2\,b_2\,A\,\sqrt{3+\alpha}\,\sinh\,\left(\frac{\sqrt{3+\alpha}}{2}\,\chi\right)\cosh\,\left(\frac{\sqrt{3+\alpha}}{2}\,\phi\right)+b_3\,A\,\sqrt{3+\alpha}\,\sinh\,\left(\sqrt{3+\alpha}\,\phi\right)\,.
\en
By integrating \eqref{ema8} and \eqref{ema9} with respect to $\phi$ and $\chi$ respectively, we obtain the Hubble parameter 
\be \label{ema12}
H(t)=\frac{A}{2} \left\{4\,b_2 \cosh^{\,2}\,\left[\log\,\left[\coth\left(\frac{3+\alpha}{2} \,A\, t\right)\right]^{\,\frac{1}{2}}\right]-(b_2-1) \left[\coth\,\left(\frac{3+\alpha}{2}At\right)+\tanh\,\left(\frac{3+\alpha}{2}At\right)\right]\right\}\,,
\ee
Equation (\ref{ema12}) makes one able to write the scale factor $a(t)$ and the EoS parameter $\omega$. Those are respectively given by

\be \label{ema13}
a(t)=\left[\cosh\,\left(\frac{3+\alpha}{2}\,A\,t\right)\right]^{\,\frac{1-b_2}{3+\alpha}}\,\left[1-\coth\,\left(\frac{3+\alpha}{2}\,A\,t\right)\right]^{-\frac{2\,b_2}{3+\alpha}}\,\left[\coth\,\left(\frac{3+\alpha}{2}\,A\,t\right)\right]^{\,\frac{b_2}{3+\alpha}}\,\left[\sinh\,\left(\frac{3+\alpha}{2}\,A\,t\right)\right]^{\frac{1-b_2}{3+\alpha}}\,,
\ee
\be\label{emax}
\omega=\frac{2 (\alpha +3)}{3\left[b_2\, \sinh \left((\alpha +3) \,A\, t\right)+\cosh \left((\alpha +3) \,A\, t\right)\right]^2}-1\,.
\ee
We can observe that the interactions between the fields $\phi$ and $\chi$ lead us to the following relations for the first two slow-roll parameters: 
\be \label{ema14}
\epsilon_1=\frac{3+\alpha }{\cosh\,\left((3+\alpha )\,A\,t\right)+b_2 \sinh^{\,2}\,\left((3+\alpha )\,A\,t\right)}\,;
\qquad
\epsilon_2 = \frac{(3+\alpha ) \left[1-\cosh\,\left(2\,(3+\alpha )\,A\,t\right)-b_2\, \sinh\,\left(2\,(3+\alpha)\,A\,t\right)\right]}{\cosh\,\left((3+\alpha )\,A\,t\right)+b_2\, \sinh^{\,2}\,\left((3+\alpha )\,A\,t\right)}\,.
\ee
\end{widetext}
Therefore, it is not possible to obtain general forms for $\epsilon_{2\,n}$ and for $\epsilon_{2\,n+1}$ when working with general values of $b_2$. Consequently we need to study the behavior of different slow-roll parameters case by case.

\begin{figure}[h!]
\vspace{1cm}
\includegraphics[{height=04cm,angle=00}]{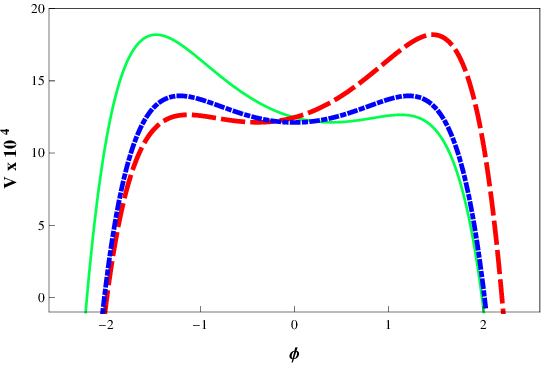}
\vspace{0.3cm}
\caption{Plots of the potential $V(\phi,\chi)$ where $A=1$, $b_2=200$, $\alpha=-2 $, $\chi=0$ for the blue (dot-dashed) curve, $\chi=-1$  for the red (dashed) curve, and $\chi=1$ for the green (solid) curve. We can note the deformation on $V$ for different values of $\chi$.}
\label{FIG1}
\end{figure}

\begin{figure}[h!]
\vspace{1cm}
\includegraphics[{height=04cm,angle=00}]{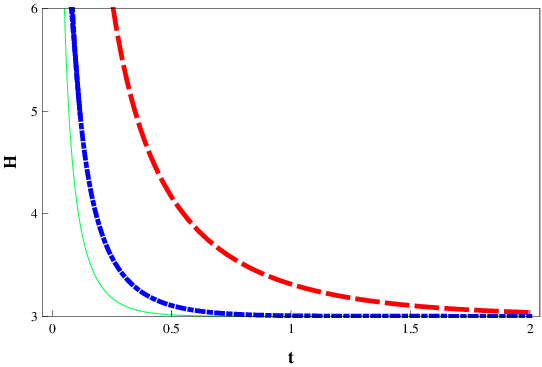}
\includegraphics[{height=04cm,angle=00}]{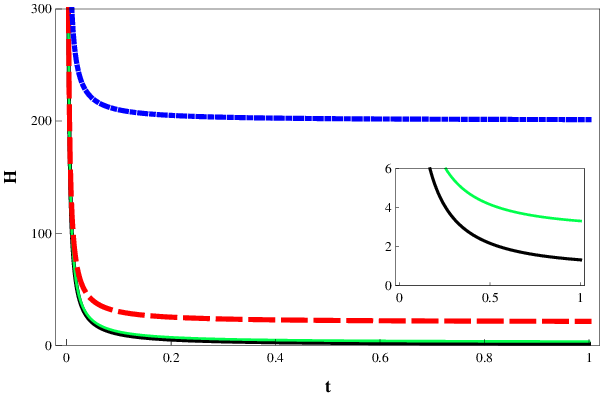}
\vspace{0.3cm}
\caption{Evolution of the Hubble parameter.}
\label{FIG2}
\end{figure}

\begin{figure}[h!]
\vspace{1cm}
\includegraphics[{height=04cm,angle=00}]{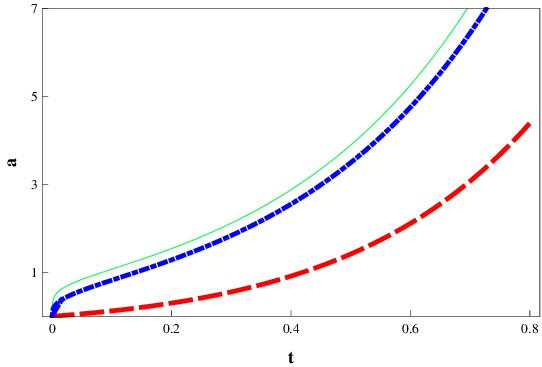}
\includegraphics[{height=04cm,angle=00}]{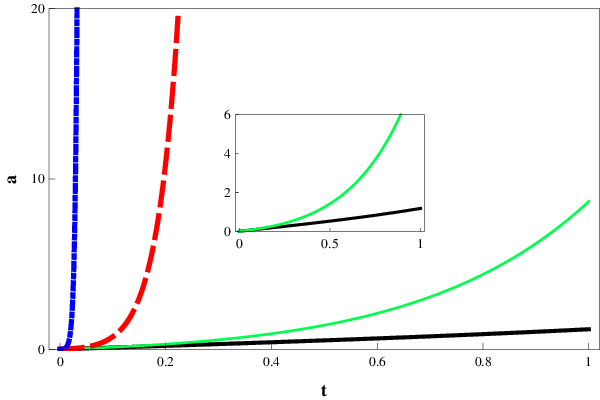}
\vspace{0.3cm}
\caption{Evolution of the scale factor.}
\label{FIG3}
\end{figure}

\begin{figure}[h!]
\vspace{1cm}
\includegraphics[{height=04cm,angle=00}]{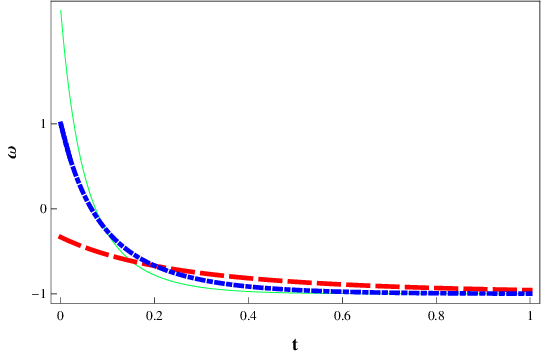}
\includegraphics[{height=04cm,angle=00}]{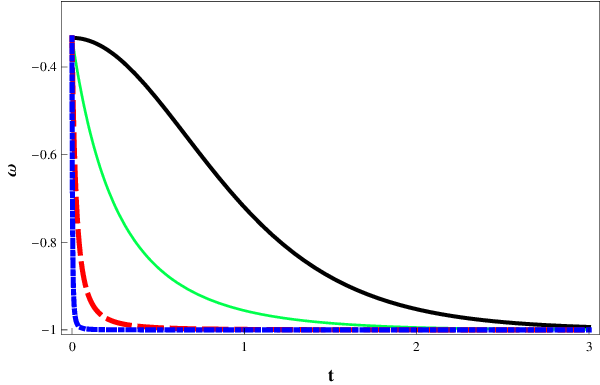}
\vspace{0.3cm}
\caption{Evolution of the equation of state parameter.}
\label{FIG4}
\end{figure}

\begin{figure}[h!]
\vspace{1cm}
\includegraphics[{height=04cm,angle=00}]{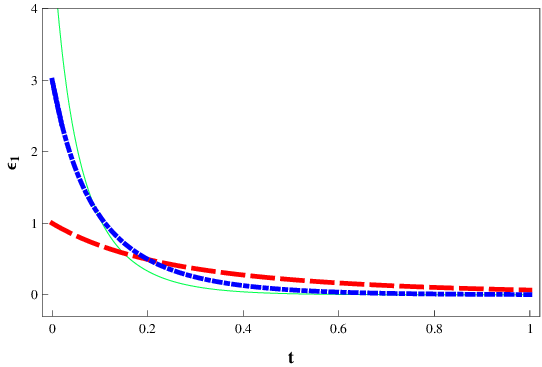}
\includegraphics[{height=04cm,angle=00}]{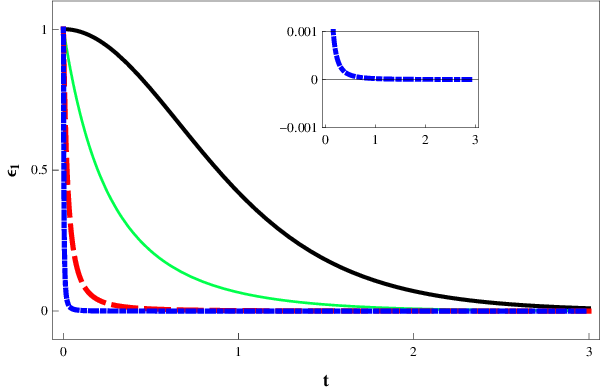}
\vspace{0.3cm}
\caption{Evolution of the slow-roll parameter $\epsilon_1(t)$.}
\label{FIG5}
\end{figure}

\begin{figure}[h!]
\vspace{1cm}
\includegraphics[{height=04cm,angle=00}]{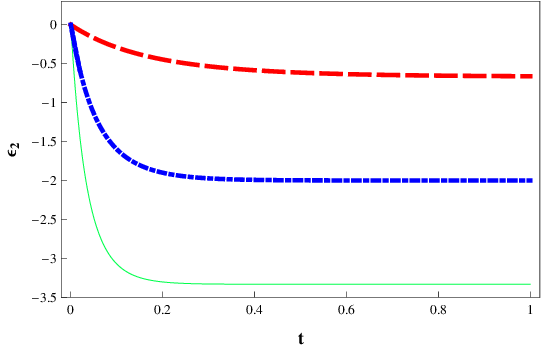}
\includegraphics[{height=04cm,angle=00}]{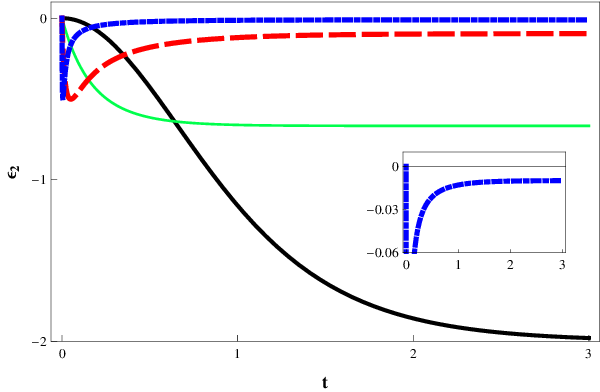}
\vspace{0.3cm}
\caption{Evolution of the slow-roll parameter $\epsilon_2(t)$.}
\label{FIG6}
\end{figure}

\begin{figure}[h!]
\vspace{1cm}
\includegraphics[{height=04cm,angle=00}]{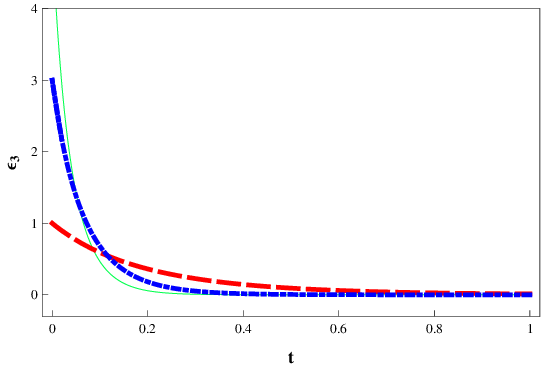}
\includegraphics[{height=04cm,angle=00}]{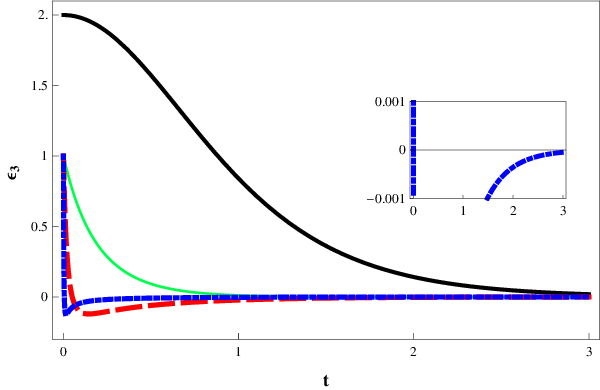}
\vspace{0.3cm}
\caption{Evolution of the slow-roll parameter $\epsilon_3(t)$.}
\label{FIG7}
\end{figure}

\begin{figure}[h!]
\vspace{1cm}
\includegraphics[{height=04cm,angle=00}]{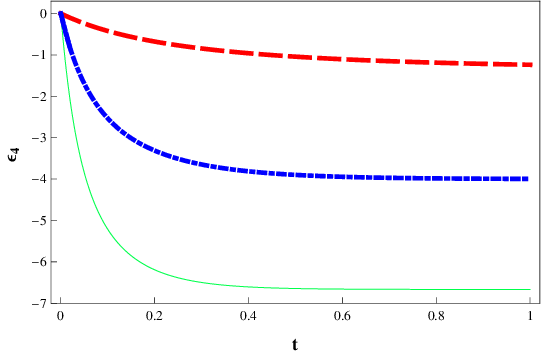}
\includegraphics[{height=04cm,angle=00}]{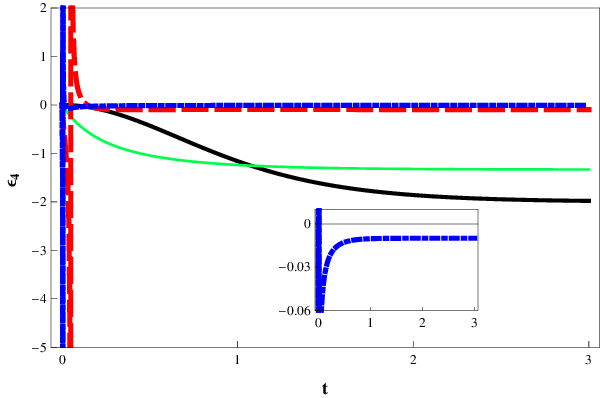}
\vspace{0.3cm}
\caption{Evolution of the slow-roll parameter $\epsilon_4(t)$.}
\label{FIG8}
\end{figure}

Below we present the behavior of the quantities obtained from the model. Fig.\ref{FIG1} characterizes the form of the potential $V(\phi,\chi)$. Figs.\ref{FIG2}-\ref{FIG4} show the evolution of the cosmological parameters $H,a$ and $\omega$ through time and Figs.\ref{FIG5}-\ref{FIG8} present the evolution of the $\epsilon_1$, $\epsilon_2$, $\epsilon_3$ and $\epsilon_4$ slow-roll parameters.

For Figs.\ref{FIG2}-\ref{FIG8}, the upper panel shows the quantity evolution by assuming $A=1$, $b_2=2$, $\alpha=0 $ for the blue (dot-dashed) curve, $\alpha=-2$  for the red (dashed) curve, and $\alpha=2$ for the green (solid) curve. In the lower frames, $A=1$, $\alpha=-2$, $b_2=200$ for the blue (dot-dashed) curve, $b_2=20$  for the red (dashed) curve, $b_2=2$ for the green (solid) curve, and $b_2=0$ for the black (solid) curve.




\section{The cosmological interpretations}\label{sec:cs}

With the purpose of evading the flatness problem, the horizon problem and the absence of magnetic monopoles, A. Guth has suggested a model of inflation \cite{guth/1981}, for which the Universe would have passed through a phase of accelerated expansion right after the Big-Bang, which lasted a small period of time. According to standard cosmology, during inflationary phase, the Hubble parameter remains constant, i.e., the scale factor grows exponentially with time as $a(t)\propto e^{H_\iota t}$, with $H_\iota$ being the Hubble parameter value at inflation era. Note that, mathematically, what would indeed make the Hubble factor $\dot{a}/a$ to be a constant is an exponential behavior for the scale factor. Let us see if these features are predicted by the solutions presented above.

The lower panel in Fig.\ref{FIG2} seems to be in agreement with an exponential expansion Universe. Firstly, note that it presents only positive values for $H$, which is required in an expanding Universe. However what should be really highlighted is the small interval of time which $H$ takes to assume constant values. Note that in all cases, the values of $H$ decrease after Big-Bang ($t=0$) and quickly enters a steady regime. On the other hand, the curves in the upper panel present a less abrupt trend to constant values of $H$. 

By analyzing the lower panel of Fig.\ref{FIG3}, it is quite clear that blue and red curves indicate a Universe which suffers an abrupt expansion in a small period of time, in total agreement with inflationary era. By making the analogy with standard model of cosmology, in which one assumes $a(t)$ to grow exponentially with time, as mentioned above, the difference in the behavior of the red and blue curves when compared to the green and black ones would lie in the values assumed for $H_\iota$.

Let us now check the lower panel in Fig.\ref{FIG4}. It is well known that the essential feature of an inflationary Universe is the acceleration of its expansion, i.e., $\ddot{a}>0$. From standard FEs, a positive $\ddot{a}$ implies $-1\leq\omega\leq-1/3$, which is precisely the range of values assumed by our solution. Once again, the parameters derived in our model describe successfully the period of inflation. Note that a similar behavior for $\omega$ was obtained in \cite{ilic/2010}.

Moreover it should be stressed that the values of $\omega$ in all curves tend to $-1$ as time passes by. In fact, that is the value of the present $\omega$ according to CMB anisotropy observations \cite{hinshaw/2013}. The present phase of acceleration our Universe is passing through, in principle, is not dissimilar to the primordial one. Therefore the agreement between the early and late-time values of $\omega$ is somehow expected.

The lower panels illustrated in Figs.\ref{FIG2}-\ref{FIG8} show that we can have different cosmological scenarios for a unique value of $\alpha$. Such a behavior generalizes the models obtained in \cite{motohashi/2014}, where the authors presented different fast-roll models only if one is dealing with different values of $\alpha$.

We can also observe that the odd slow-roll parameters are always negligible as time passes by. The even slow-roll parameters can be of order unity for some set of choices of the constants $A$, $b_2$ and $\alpha$. However, Fig.\ref{FIG6} and Fig.\ref{FIG8} unveil the dependence of the even slow-roll parameters with the constant $b_2$. It is clear that these last parameters may be smaller as we take bigger values of $b_2$, which is the constant that is effectively coupling the fields $\phi$ and $\chi$ in our model.

\section{Final Remarks and Perspectives}\label{sec:d}

In this work, we have presented solutions to a two scalar fields model in fast-roll regime. In order to do so, we have applied the method introduced in \cite{bglm}. The results presented in Subsection \ref{ss:ema} are observationally acceptable in a Universe undergoing a phase of accelerated expansion.

Since our intention was to obtain a physically well-behaved model, it is worth stress that the solutions (\ref{ema12})-(\ref{emax}), related to the Hubble, acceleration and EoS parameters, successfully undergo a dimensional analysis, as follows. Let us, firstly, analyze the Hubble parameter. Since $H\propto \tau_H$, with $\tau_H$ being the Hubble time, $H$ should have its units given in $s^{-1}$. From Eq.(\ref{ema12}), note that such a property may be satisfied if the constants $b_2$ and $\alpha$ are dimensionless and $A$ has units given in $s^{-1}$. If these features are employed in (\ref{ema13})-(\ref{emax}), the scale factor and EoS parameter happen to be dimensionless, as they, in fact, should be.

As mentioned in Section \ref{sec:cs}, the Hubble parameter evolution presented in the lower panel of Fig.\ref{FIG2} is in agreement with an inflationary Universe. 

Note that Eq.(\ref{ema13}) presents a scale factor evolution quite different from those obtained in the so-called ``power-law inflation" (check \cite{lucchin/1985,lyth/1992}), for which $a$ grows like $t^{p}$, with $p>1$. Those scenarios arise, for example, when one has a single scalar field with an exponential potential. Anyhow, the scale factor evolution derived for our model and presented in Fig.\ref{FIG3}, likewise describes an inflationary scenario, specially through the lower panel blue and red curves.

The EoS evolution in Fig.\ref{FIG4} also successfully describes the inflationary era. Specifically, its lower panel restricts the values of $\omega$ to those for which it is possible to obtain an accelerated expansion for the Universe according to standard FEs, i.e. $-1\leq\omega\leq-1/3$. Anyway, all curves for $\omega$ have its values tending to $-1$ as time passes by, which is indispensable in an inflationary Universe.

Moreover, this two-field description gives us some control concerning the slow-roll parameters as we see in Figs.\ref{FIG5}-\ref{FIG8}. Therefore the fact that the even slow-roll parameters may be less than order unity for some set of constants, together with the behavior of the another cosmological parameters, strengthens our hypothesis that the two-field fast-roll model can describe an inflationary era of the Universe. We also point that the procedures here presented can be applied in other interesting theories of gravity such as $f(Q)$, $f(R,T)$ and $f(Q,T)$ gravity, as one can see in Refs. \cite{harko, ms_16, Jimenez/2018, Mandal/2020, Yixin/2019, Simran/2020}. The study of the compatibility of the fast-roll inflation for these theories of gravity is a interesting challenge to tackle, specially when one deals with hybrid inflation, and we hope to report on such subjects in near future. 

\acknowledgments P.H.R.S. Moraes would like to thank CAPES for financial support. JRLS would like to thank CNPq (Grant no. 420479/2018-0), CAPES, and PRONEX/CNPq/FAPESQ-PB (Grant nos. 165/2018, and 0015/2019) for financial support.

The authors inform that the discussions carried in this article are also presented in the preprint titled ``Fast-roll solutions from two scalar fields inflation", which can be access through the link https://arxiv.org/abs/1504.07204

\pagebreak


\end{document}